\documentclass[twocolumn,aps,prx,showpacs,superscriptaddress]{revtex4-2}

\usepackage{amsmath}
\usepackage{amsfonts}
\usepackage{amssymb}
\usepackage{amsmath}
\usepackage{amsthm}
\usepackage{mathrsfs}
\usepackage{graphicx}
\usepackage{color}
\usepackage{amsfonts}
\usepackage{wrapfig}
\usepackage{cases}
\usepackage{framed}
\usepackage{float}
\usepackage[colorlinks,citecolor=blue]{hyperref}
\usepackage{multirow}

\usepackage{bigints}
\usepackage[mathabx]{varint}

\begin{document}



\title{Quantum Theory for Condensation Phase of Nonequilibrium Bosons: Graph topology of flux network and fluctuations}

\author{Feihong Liu}
\affiliation{Department of Materials Science \& Engineering, City University of Hong Kong, Kowloon, Hong Kong SAR}

\author{Chase Slowey}
\affiliation{Department of Chemistry, University of North Carolina at Chapel Hill, Chapel Hill, North Carolina 27599, United States}

\author{Xuanhua Wang}
\affiliation{School of Arts and Sciences, Fuyao University of Science and Technology, Fuzhou, Fujian 350122, China}

\author{Dangyuan Lei}
\affiliation{Department of Materials Science \& Engineering, City University of Hong Kong, Kowloon, Hong Kong SAR}
\affiliation{Department of Physics, City University of Hong Kong, Kowloon, Hong Kong SAR}

\author{Jeremie Torres}
\affiliation{Laboratoire Charles Coulomb, University of Montpellier CNRS, Montpellier, France}
\affiliation{Quantum Biology Lab, Howard University, Washington D.C. 20059, United States}

\author{Zhiyue Lu}
\email{zhiyuelu@unc.edu}
\affiliation{Department of Chemistry, University of North Carolina at Chapel Hill, Chapel Hill, North Carolina 27599, United States}

\author{Zhedong Zhang}
\email{zzhan26@cityu.edu.hk}
\affiliation{Department of Physics, City University of Hong Kong, Kowloon, Hong Kong SAR}
\affiliation{Shenzhen Research Institute, City University of Hong Kong, Shenzhen, Guangdong 518057, China}

\date{\today}

\begin{abstract}
Condensation of nonequilibrium bosonic systems subject to energy pump and dissipation are investigated, where strong coherence emerges far from equilibrium. A quantum theory is developed to capture such a nonequilibrium nature, yielding a certain graphic structure arising from the detailed-balance breaking. The results show a network of probability curl fluxes--resulting from quantum fluctuation of energy--that reveals a graph topology. The winding number associated with the flux network is thus identified as a new order parameter for the nonequilibrium phase transition, not attainable by the symmetry breaking. 
Our work demonstrates a new nonequilibrium phase of matter, revealing an alternative topological phase transition towards robust coherence order.
\end{abstract}

\maketitle

\section{Introduction}
Driven-dissipative systems, as typical nonequilibrium process, have drawn great attention in physical and chemical research \cite{Jeremie_SA2025,Lagoudakis_NP2019,Xu_PRR2025,Amo_Nature2009,Xiong_ACR2023,Liang_PRL2024,Wang_NatMat2025}. As a prominent member of the driven-dissipative components, Bose-Einstein condensation (BEC) of atoms and Fr\"ohlich condensation of phonons are presented by collective motion and giant dipole moments  \cite{Frohlich_IJQC1968,Preto_JBP2017,Wu_JTB1978,Lundholm_SD2015}. These activate long-range forces, enabling lasing gain and coherent energy transfer observed in light-harvesting antennas  \cite{Jeremie_SA2025}. Provided the synergy of energy pump and dissipation, a collective order may appear--as proposed by Fr\"ohlich--that induces dissipative structures \cite{Frohlich_IJQC1968,Prigogine_Chapter1973,Popp_CBP1984,Wu_JBP1981}. Such an emerging order renders nonequilibrium phase transitions, offering a unified understanding of collective effects in complex materials, e.g., giant dipole in low-energy range, many-particle phases and cognitive function \cite{Tuszynski_PRA1984,Moessner_ARCMP2022,Hameroff_PTRSA1998,Song_SA2025,Schneider_Nature2013,Feist_NatCommun2016,Tuszynski_PRE2002}. Although the driven-dissipative dynamics has been explored broadly, the coherence in far-from-equilibrium regime still remains elusive.

So far, the driven-dissipative phases have been observed in various systems including molecules, polymers and semiconductors  \cite{Frohlich_IJQC1968,Reimers_PNAS2009,Zhang_PRL2019,Kasprzak_Nature2006,Xiong_NM2021,Snoke_PRL2017}. Nardecchia, et al., observed the phonon condensation in model proteins at 0.3THz using the THz absorption \cite{Nardecchia_PRX2018}. In this vein, the advancements of the X-ray spectroscopic and crystallographic techniques enabled a real-time snapshot of the structural oscillation which are long-lived in protein crystals \cite{Lundholm_SD2015,Miao_Nature1999}, orders of magnitude longer than that for thermalization process \cite{Turton_NC2014}. The optical gain in conjunction with the long-range electric forces, as measured in recent experiments, indicates the collectivity of low-energy excitations \cite{Jeremie_SA2025,Jeremie_SA2022}. By extending the parameters to optical range, the driven-dissipative phases can occur in cavity polaritons, as observed in pumped semiconductors and magnons that constitute the long-range order, e.g., polariton condensates   \cite{Kasprzak_Nature2006,Deng_Science2002,Esmann_2024,Mahrt_NM2014,Yamamoto_RMP2010,Zhou_NC2019}. Despite these achievements, the fluctuations--closely related to the coherence--of nonequilibrium systems are still an open issue. The irreversible nature may lead to unusual fluctuations, indicating nonthermal distributions. A full counting statistics deviating from the Boltzmann’s law still lacks, although several indications were actively discussed before \cite{Yamamoto_RMP2010,Schwendimann_PRB2008,Zhang_PRB2022,Wang_PRB2022,Schneider_PRL2018,Laussy_PRL2004,Liu_PRL2023,Ostrovskaya_NC2020}. These call for a different mechanism for understanding bosonic collectivity from the BEC one, underlying the essential of exceeding the scope of symmetry breaking \cite{Yamamoto_NP2014,Butov_Nature2007,Butov_NPht2012,Pledran_NPht2012,Ostrovskaya_NC2018,Yang_RMP1962,Leggett_RMP2001,Jiang_PRA2014,Sieberer_RPP2016}.

In this article, we propose a nonequilibrium condensation of spinless bosons (NCB), which possess a graph topology. The NCB is formed with energy drive above a threshold. A quantum theory beyond the paradigm of U(1) symmetry breaking is developed. This evidences a different class of matter phases from the atomic BECs. Our model reveals a strong correlation with total particle number, which breaks the detailed balance. A probability flux network of 2D lattice graphs emerges with certain topological structures. We elaborate on the curl nature of the flux network and identify the winding number as a new order parameter for such nonequilibrium phase transition. The topological essence, as shown here, provides a measure of the detailed-balance breaking, indicating the power of network analysis. Our model also provides experimental signatures associated with the flux network, i.e., coherent oscillations in time-resolved photon-coincidence counting spectra.

\section{Driven-dissipative model of phonons}

The vibrations in molecules, e.g., vibrations of J/H aggregates and DNA backbones, are surrounded by dense mediums (like solvent or water) that act as a thermal environment. This results in the channels of energy dissipation and internal conversion. The latter is responsible for the energy redistribution amongst the vibrations, thus generating the nonlinearity. Besides, the vibrations are driven by an external energy pump.

\begin{figure}[t]
\centering
\includegraphics[scale=0.335]{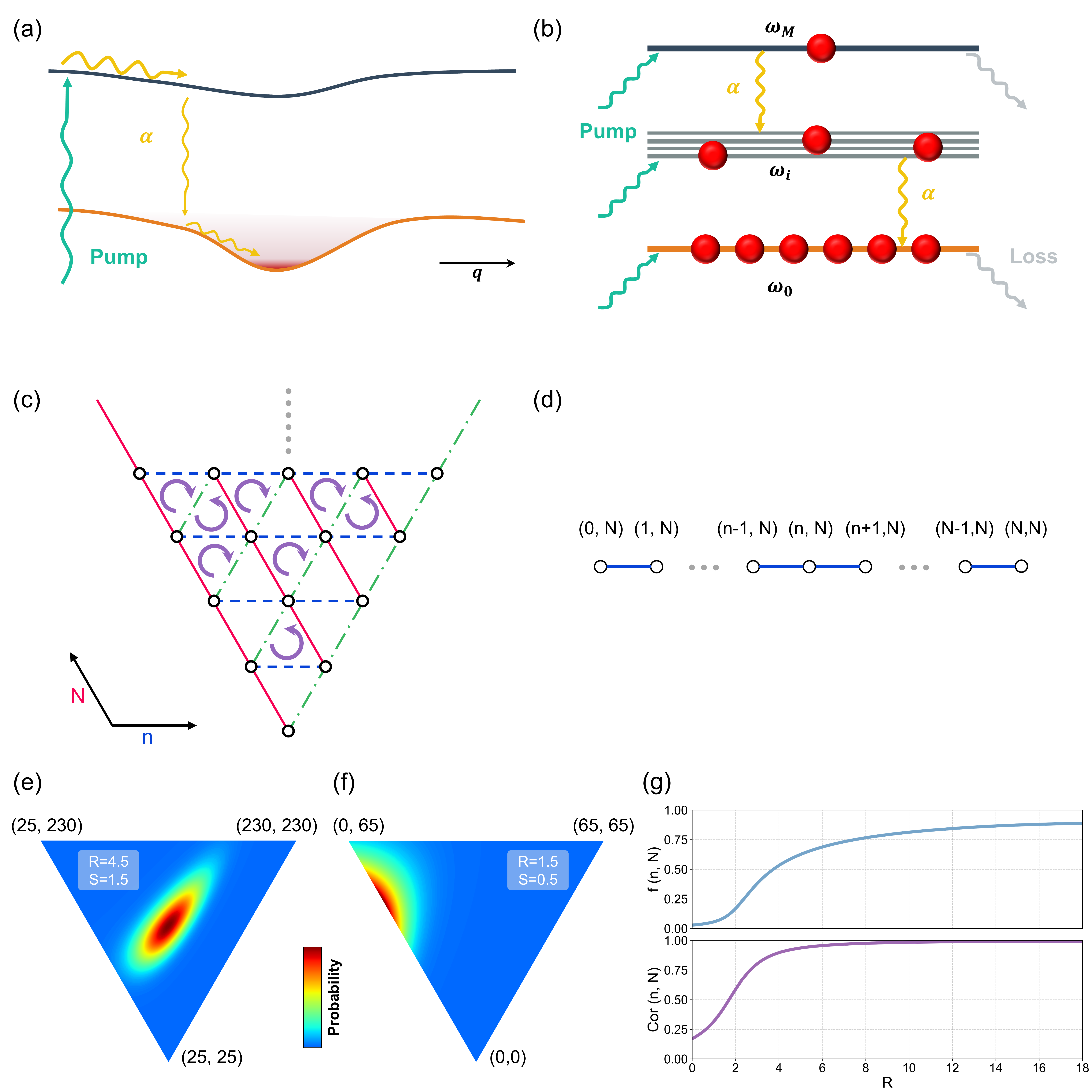}
\caption{(a) Schematic illustration of phonon dispersion and relaxation towards the lowest-energy mode. (b) Level structure for bosons subject to external energy pump and dissipation. (c) 2D hexagonal-grid graph for nonequilibrium bosons mapped from Eq.(\ref{EOMP}), with nonvanishing net currents. (d) 1D tree graph for the BEC phase where the net currents vanish. (e) Steady-state number distribution $P_{n,N}$ when pump is above the threshold; (f) $P_{n,N}$ when pump is below the threshold. (g,up) Condensation fraction $f$ against pump rate $[R = 3S]$; (g,down) Pearson correlation between $n$ and $N$ against pump rate $[R = 3S]$.}
\label{Schematic}
\end{figure}

For a neat picture, we adopt a generic model of phonons with a sandwich structure of energy levels that contains $M$ modes depicted in Fig.\ref{Schematic}(a). The lowest mode has the energy $\omega_0$, whereby the excited modes are densely distributed that yields a high density of states within a narrow bandwidth of energy. The free Hamiltonian is $H_0 = \sum_{i=0}^M \omega_i \eta_i^{\dagger} \eta_i$; $\omega_1,\omega_2,...,\omega_M$ denote the energies of excited modes. $\eta_i$ and $\eta_i^{\dagger}$ are the respective bosonic annihilation and creation operators, i.e., $[\eta_i,\eta_j^{\dagger}] = \delta_{ij}$. With the effects of external pump and environment, the density matrix obeys the dynamical equation
\begin{equation}
  \begin{split}
    \dot{\rho} = -{\rm i}[H_0,\rho] + \big(\hat{\text{W}}_{\ell} + \hat{\text{W}}_{n\ell} \big)\rho
  \end{split}
\label{EOM}
\end{equation}
where the superoperators are:  $\hat{\text{W}}_{\ell} \rho = \sum_x \gamma_x (\pi_x^{\dagger} \rho \pi_x - \rho \pi_x \pi_x^{\dagger}) + \text{h.c.}$ with $\pi_x = \{\eta_i, \eta_i^{\dagger}\}$ from pump \& dissipation channels (linear); $\hat{\text{W}}_{n\ell} \rho = \sum_{i,j} \chi_{ij} (\eta_i^{\dagger} \eta_j \rho \eta_j^{\dagger} \eta_i - \rho \eta_j^{\dagger} \eta_i \eta_i^{\dagger} \eta_j) + \text{h.c.}$ from the internal conversion channels (nonlinear) [Appendix A].

\section{Reduced dynamics of the lowest mode}
Defining the reduced density matrix for the $\omega_0$ mode, $\sigma_{n,N;n+\delta,N+\delta} = \sum'_{\{n_k\}}\langle n; \{n_k\}|\rho|n+\delta; \{n_k\} \rangle$ 
with $N$ as the total particle number and $\{n_k\}\equiv n_1,...,n_{M}$ where the $\sum'_{\{n_k\}}$ is subject to $n+n_1+\cdots+n_M = N$, the reduced equation of motion (rEOM) can be derived from Eq.(\ref{EOM}) [Appendix B]. Proceeding with these lines we find ($P_{n,N}\equiv \sigma_{n,N;n,N}$) \cite{OF}
\begin{equation}
  \begin{split}
     \dot{P}_{n,N} = & + n\big[R P_{n-1,N-1} - (R+1)P_{n,N}\big] \\[0.15cm]
    & - (n+1)\big[R P_{n,N} - (R+1)P_{n+1,N+1}\big]  \\[0.15cm]
    & + \big[S {\cal A}_{n,N-1}P_{n,N-1} - (S+1){\cal B}_{n,N}P_{n,N}\big] \\[0.15cm]
    & - \big[S {\cal A}_{n,N}P_{n,N} - (S+1){\cal B}_{n,N+1}P_{n,N+1}\big] \\[0.15cm]
    & + \alpha n \big[{\cal K}_{n-1,N}P_{n-1,N} - {\cal H}_{n,N}P_{n,N}\big] \\[0.15cm]
    & - \alpha (n+1) \big[{\cal K}_{n,N}P_{n,N} - {\cal H}_{n+1,N}P_{n+1,N}\big]
  \end{split}
  \label{EOMP}
\end{equation}
with parameters ${\cal A}_{n,N}$, ${\cal B}_{n,N}$, ${\cal K}_{n,N}$, ${\cal H}_{n,N}$ which can be either obtained from phenomenological way or microscopic model \cite{ABKH}. $R$, $S$, $\alpha$ are the respective rates of energy pump at $\omega_0$ mode, higher modes, the nonradiative transition rate between the phonon modes; all have been rescaled by the radiative decay rate. We evaluate ${\cal A}_{n,N}$, ${\cal B}_{n,N}$, ${\cal K}_{n,N}$, ${\cal H}_{n,N}$ in varying degrees of rigor. One of the most illuminating is to note that for solvent environments the thermal factor $\bar{n}_{j0} = [e^{(\omega_j-\omega_0)/T} - 1]^{-1}$, as they appear in ${\cal K}_{n,N}$ and ${\cal H}_{n,N}$, can be assumed as uniform, i.e., $\bar{n}_{j0}\approx \bar{n}$. Then the parameters are
\begin{subequations}
    \begin{align}
        & {\cal A}_{n,N} = N + M - n, \quad {\cal B}_{n,N} = N - n \label{AB} \\[0.15cm] 
        & {\cal K}_{n,N} = \left(\bar{n} + 1\right)(N-n),\  {\cal H}_{n,N} = \bar{n} (N+M-n). \label{KH}
    \end{align}
\end{subequations}

Eq.(\ref{EOMP}) can be recast into the form in Liouville space 
\begin{equation}
  |\dot{P}\rangle = W |P\rangle,\ \ |P\rangle =(P_{0,0},P_{0,1},\cdots,P_{n,N},\cdots)^{\text{T}}.
\end{equation}
$W_{n',N';n,N}P_{n,N}$ accounts for the rate of forward transition $(n,N)\rightarrow (n',N')$, and then the rate of backward transition $(n',N')\rightarrow (n,N)$ follows, i.e., $W_{n,N;n',N'}P_{n',N'}$. The two rates are not equal normally, and form a probability network as depicted in Fig.\ref{Schematic}(c). The detailed-balance breaking is therefore revealed.


\section{Condensation of nonequilibrium bosons}

From Eq.(\ref{EOMP}) one can find the equations for the means of $n$ and $N$, i.e.,
\begin{subequations}
  \begin{align}
    & \langle \dot{n}\rangle = \alpha \langle (N-C)n \rangle - \alpha \langle n^2\rangle + \alpha (\bar{n}+1) \langle N\rangle + R, \label{nd} \\[0.15cm]
    & \langle \dot{N}\rangle = R + MS - \langle N\rangle \label{Nd}
  \end{align}
\end{subequations}
with $C=(M+1)\bar{n} + 1 + \alpha^{-1}$. 
Eqs.(\ref{nd},\ref{Nd}) evidence a pump threshold $R_{\text{c}}, S_{\text{c}}$, above which an inflation of the $\omega_0$ occupation number appears--a condensation at $\omega_0$ state. For a back-of-the-envelop understanding, one can presumably neglect the fluctuation of $N$, i.e., $\langle n N\rangle \approx \langle n\rangle N$. This gives an estimation $R_{\text{c}}+MS_{\text{c}}=C$.

\begin{figure}[t]
\centering
\includegraphics[scale=0.45]{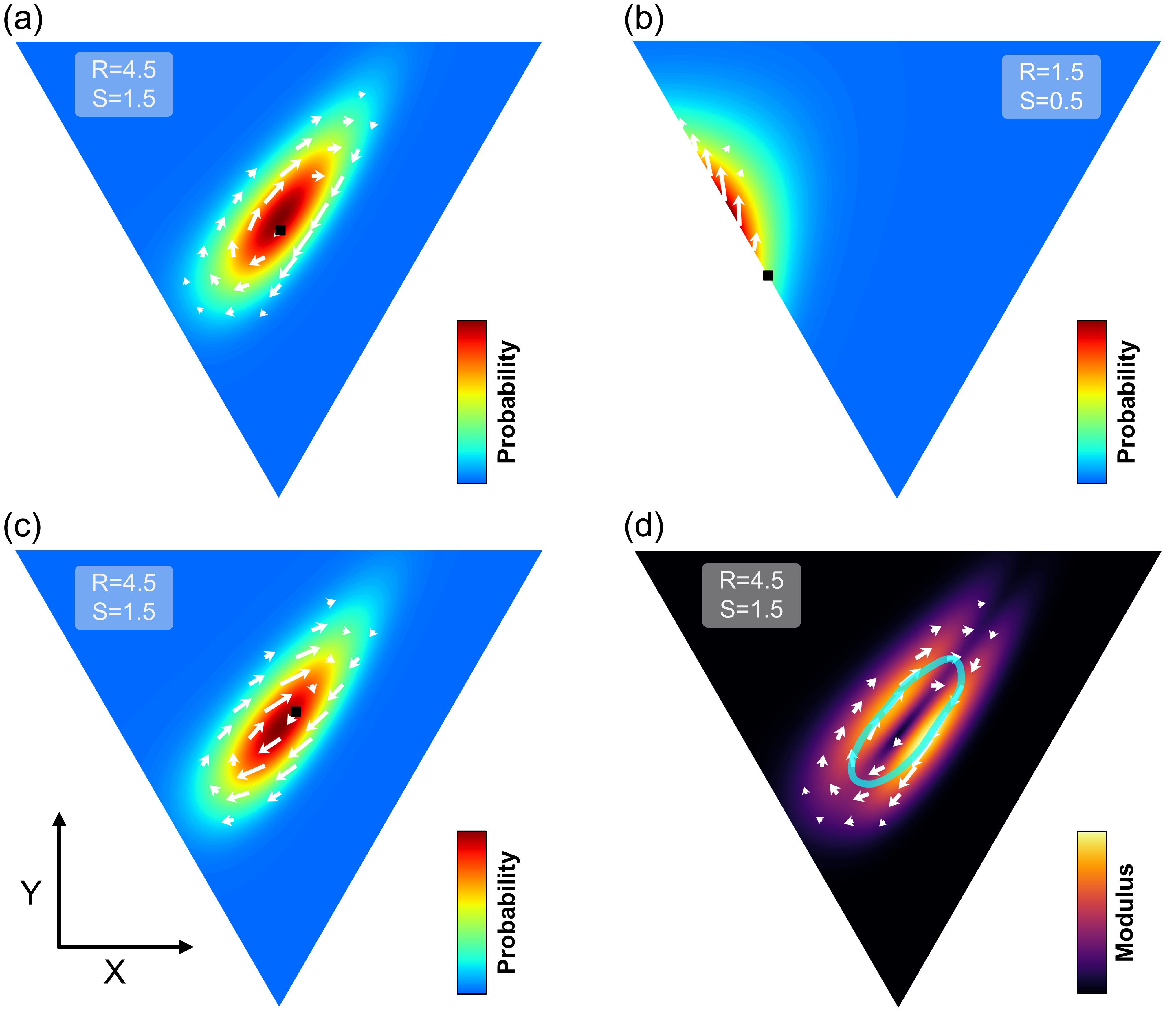}
\caption{Curl flux (white arrows) ${\rm J}_{\rm c}$ for (a) NCB phase in above-threshold regime, (b) thermal phase in below-threshold regime, where the flux is calculated from Eq.(\ref{Jc}). Obviously, no loops are presented in thermal phase. (c) Current network (white arrows) obtained from Eq.(\ref{Jnet}), showing the curl nature consistent with (a). (d) Modulus $|{\rm J}_{\rm c}|$ revealing a summit-crater landscape with a ring ridge; green loop on the ridge locates the maximum curl flux. Black square spot denotes the singularity $z_{\rm m}$ of ${\rm J}_{\rm c}$.}
\label{flux}
\end{figure}

The means of $n$ and $N$ can be found from Eq.(\ref{EOMP}). This enables the fraction $f = \frac{\langle n\rangle}{\langle N\rangle}$ that reveals the condensation transition in an explicit way, once being aware of drastic fluctuations around critical point. We plot the results in Fig.\ref{Schematic}(g,up), depicting a sharp increase towards a high $f$, when above a pump threshold $R_{\text{c}}, S_{\text{c}}$.

Nevertheless, one should note that the $n,N$ fluctuations are strongly correlated with each other, when above $R_{\text{c}}, S_{\text{c}}$. Such an insight underlines the essential of an advanced theory beyond the present understanding. For instance, the curl flux arising from the $n,N$ fluctuations may lead to oscillation that is absent from the mean-field treatment.


\section{Curl flux network and detailed balance violation}

Eq.(\ref{EOMP}) can be solved at steady state $\dot{P}_{n,N} = 0$ for the distribution $P_{n,N}$, 
depicted in Fig.\ref{Schematic}(e,f). With higher pump power, $P_{n,N}$ reveals two insights: (1) a transition from thermal to coherent statistics; (2) a stronger $n$-$N$ correlation once the condensates are formed [see Fig.\ref{Schematic}(g,down)]. The nonequilibrium nature is thus inferred, signifying the fluctuation of $N$.

The transition rates $W_{n',N';n,N}$ between neighboring sites can be mapped to bi-directional edge graphs of hexagonal grids, shown in Fig.\ref{Schematic}(c). The net currents on edges read (Appendix C, D)
\begin{equation}
  J_{(n,N)\rightarrow (n',N')} = W_{n',N';n,N}P_{n,N} - W_{n,N;n',N'}P_{n',N'}
\label{Jnet}
\end{equation}
and $W_{n,N;n-1,N-1} = n R$, $W_{n-1,N-1;n,N} = n(R+1)$, etc.. Using Eq.(\ref{Jnet}), one recasts Eq.(\ref{EOMP}) into
\begin{equation}
  \begin{split}
   \dot{P}_{n,N} =  & \  J_{(n-1,N-1)\rightarrow (n,N)} - J_{(n,N)\rightarrow (n+1,N+1)} \\[0.15cm]
   & + J_{(n,N-1)\rightarrow (n,N)} - J_{(n,N)\rightarrow (n,N+1)}\\[0.15cm]
   & + J_{(n-1,N)\rightarrow (n,N)} - J_{(n,N)\rightarrow (n+1,N)}.
  \end{split}
\label{EOMPJ}
\end{equation}
Eq.(\ref{EOMP}) thus forms certain tree connections, depicted in Fig.\ref{Schematic}(c),  giving the topology of the graph \cite{Hill_JTB1966,Hill_PNAS1975}. Eq.(\ref{EOMPJ}) can form a flux network on top of the edge graph. The flux network has been studied in classical stochastic processes \cite{Schnakenberg_RMP1976,Wang_AP2015,Wang_PNAS2008,Qian_ARPC2007,Chamberlin_PNAS1998}. 
For quantum systems, the concept of flux network was developed in recent progresses \cite{Zhang_JCP2014,Zhang_NJP2015,Ren_PRL2022}.

$J_{(n,N)\rightarrow (n',N')}$ measures how far the system deviates from the equilibrium, thus breaking the detailed balance. When the condensates emerge, $J$s on the graph consist of nonzero loop currents at steady state. To see this closely, we essentially apply the affinity as follows.

The elementary triangles $\triangledown,\vartriangle$ are the building blocks for our hexagonal-grid graph. For a closed trajectory along $\triangledown$, Eq.(\ref{EOMP}) enables the affinity $\Phi_{\triangledown} = \ln (\Pi_{\triangledown^+}/\Pi_{\triangledown^-})$; $\Pi_{\triangledown^{+(-)}}$ is the product of the transition rates along the clockwise (counter-clockwise) direction. This gives
\begin{equation}
  \Phi_{\triangledown} = \frac{(R+1)S}{R(S+1)} \left(1 + \frac{1}{\bar{n}}\right),\quad \Phi_{\vartriangle} = -\Phi_{\triangledown}
\label{AF}
\end{equation}
so that $\Phi_{\triangledown}$, $\Phi_{\vartriangle} \neq 0$ (Appendix C). Therefore $J_{(n,N)\rightarrow (n',N')} \neq 0$ yielding the detailed balance breaking.

For a fixed $N$ when pump off, Eq.(\ref{EOMP}) reduces to 
\begin{equation}
  \dot{P}_{n,N} = J_{(n-1,N)\rightarrow (n,N)} - J_{(n,N)\rightarrow (n+1,N)}.
\end{equation}
This generates a 1D tree graph, as depicted in Fig.\ref{Schematic}(d), resulting in zero net currents at steady state. This describes the BEC phase \cite{Scully_PRL1999} (Appendix F).


It elucidates the nonequilibrium nature of the NCB, indicating a graph texture readily distinct from the BEC. This may lead to a new order parameter based on graph topology.

\section{Off-diagonal long-range order}
The reduced density matrix defined in Eq.(\ref{EOMP}) is of a standard form for a state with ODLRO. In particular, it has $\text{ODLRO} = \lim_{|m-n|\rightarrow\infty} \langle b_m^{\dagger} b_n \rangle$ where $b_n$ is the operator for the vibrations at local site. Using $\eta$ operators, one can calculate from Eq.(\ref{EOMP}) that 
\begin{equation}
    \text{ODLRO} = \frac{\langle n\rangle}{M} e^{{\rm i}\phi} + \cdots.
\end{equation}

It is known that the ODLRO is an explicit expression of the global U(1)-symmetry breaking \cite{Yang_RMP1962}. The ODLRO also exists in BEC and superfluid phases, for which we will skip the details to avoid redundancy \cite{Leggett_RMP2001,Jiang_PRA2014,Sieberer_RPP2016}.

\section{Graph topology and order parameter for condensates}
The analysis can proceed for a clear form of curl fluxes, through the continuous limit, i.e., with a large volume $V$. This defines two variables in a hexagonal frame, i.e., $x=n/V,\ y = N/V$ and $x,y$ become continuous as $V\rightarrow \infty$. In a Cartesian frame $(X,Y)$ such that 
\begin{equation}
  X = x - \frac{y}{2},\quad Y = \frac{\sqrt{3} y}{2},
\end{equation}
one has the asymptotic expansion
\begin{subequations}
 \begin{align}
  & P_{n\pm 1,N\pm 1} \rightarrow e^{\pm \hat{\mathscr{D}_+}/V}P(X,Y), \\[0.15cm]
  & P_{n\pm 1,N} \rightarrow e^{\pm \partial_X /V} P(X,Y), \\[0.15cm]
  & P_{n,N\pm 1} \rightarrow e^{\mp \hat{\mathscr{D}}_-/V} P(X,Y),
 \end{align}
\end{subequations}
with $\hat{\mathscr{D}}_{\pm} = \frac{1}{2}(\partial_X \pm \sqrt{3}\partial_Y)$ (Appendix C). This reforms Eq.(\ref{EOMPJ}) into a partial differential equation, after some algebra, i.e.,
\begin{equation}
    \partial_t P = -\nabla \cdot \left[{\bf F}P - 
    \nabla\cdot ({\bf D}P)\right]
\label{FPE}
\end{equation}
where two matrices are
\begin{subequations}
    \begin{align}
        & {\bf F}
     = - \begin{pmatrix}
    \frac{1}{2} \left[x - \frac{R}{V} + q \right] - \alpha' u(x) \\[0.2cm]
    \frac{\sqrt{3}}{2} \left[x - \frac{R}{V} - q \right]
    \end{pmatrix} \\[0.35cm]
    & {\bf D} = \frac{1}{8V}
    \begin{pmatrix}
    \left[r(x) + p\right] + 4\alpha' v(x) & \sqrt{3} \left[r(x) - p\right]\\[0.2cm]
    \sqrt{3} \left[r(x) - p\right] & 3 \left[r(x) + p\right]
    \end{pmatrix}
\end{align}
\end{subequations}
with $\alpha'=\alpha V$, $r(x) = (2R+1)x + \frac{R+1}{V},\ u(x) = \left(x + \frac{1}{V}\right) \frac{{\cal K}}{V} - x \frac{{\cal H}}{V},\ v(x) = x \frac{{\cal K}}{V} + \left(x + \frac{1}{V}\right)\frac{{\cal H}}{V}$ and $ q = S \frac{{\cal A}}{V} - (S+1)\frac{{\cal B}}{V} ,\ p = S \frac{{\cal A}}{V} + (S+1)\frac{{\cal B}}{V}$ (Appendix C).

Eq.(\ref{FPE}) defines the probability curl flux
\begin{equation}
    {\bf J}(X,Y) = {\bf F}P - \nabla\cdot ({\bf D}P).
\label{Jxy}
\end{equation}
Eq.(\ref{Jxy}) may possess nontrivial topology arising from the detailed balance breaking. To see this clearly, we extend ${\bf J}$ onto complex domain, i.e., $(X,Y)\rightarrow z\equiv X + \mathrm{i} Y$
\begin{equation}
    \text{J}_{\text{c}}(z,z^*) = \text{J}_X - \mathrm{i} \text{J}_Y.
\label{Jc}
\end{equation}

$\text{J}_{\text{c}}$ is not entirely analytic, due to the cycling nature of ${\bf J}$ at steady state. Thus
\begin{equation}
  \bigointssss \text{J}_{\text{c}}(z,z^*) \text{d}z \neq 0,\quad {\rm J}_{{\rm c}}(z_{{\rm m}},z_{{\rm m}}^*) = 0
\label{intJc}
\end{equation}
when enclosing a point $z_{{\rm m}}$ (Appendix E). 



Eq.(\ref{intJc}) indicates that $\text{J}_{\text{c}}$'s phase is ill-defined at $z_{\mathrm{m}}$. The curl flux should exhibit a vortex, yielding a  topological structure.

The topological structure of the curl fluxes can be clarified by the homotopy group. $\text{J}_{\text{c}}$ reveals a cycling nature when having the condensates--as a result of the divergence free and homogeneous boundaries shown in Fig.\ref{flux}(a)--which forms a U(1) space. This enables a mapping: $\text{U}(1)\rightarrow S_{\phi}^1$ where $S_{\phi}^1=\phi\in [0,2\pi)$ is a circle. Such a mapping can be characterized by integer winding numbers, i.e., $e^{\text{i}n\theta},\ 0\leqslant \theta < 2\pi$ \cite{Abelian}. 
One can denote this fact symbolically as, in terms of the fundamental group, $\pi_1[\text{U}(1)]=\{0,\pm 1, \pm 2,...\}$.

\begin{table}[t]
  \caption{Symmetry and topology of the NCB, BEC and thermal phases.}
  \label{T1}
  
\begin{ruledtabular}
\begin{tabular}{ccccc}
  {} & {} {} {} ODLRO & {} {} {} Symmetry & {} {} {} Homotopy & {} {} {} Winding \tabularnewline
  \hline
  NCB & {} {} {} Yes & {} {} {} $\times$ & {} {} {} $\mathbb{Z}$ & {} {} {} $\pm 1$ \tabularnewline
  Thermal & {} {} {} No & {} {} {}  U(1) & {} {} {} $\times$ & {} {} {} 0 \tabularnewline
  BEC & {} {} {} Yes & {} {} {} $\times$ & {} {} {} $\times$ & {} {} {} 0 \tabularnewline
\end{tabular}
\end{ruledtabular}

\end{table}

In contrast, a bundle of open lines (denoted by $\text{e}$) are observed at the below-threshold regime, corresponding to the thermal phase. The BEC phase has vanishing flux, generating a null space $\varnothing$. Therefore an exact mapping is established, i.e.,
\begin{equation}
  \begin{split}
    & \text{NC}: \pi_1 [\text{U}(1)] = \mathbb{Z};\quad \mathbb{Z} = \{0,\pm 1,\pm 2,...\} \\[0.15cm]
    & \text{Thermal}: \pi_1[\text{e}] = 0,\quad \text{BEC}: \pi_1[\varnothing] = 0.
  \end{split}
\label{FG}
\end{equation}

Eq.(\ref{FG}) indicates a graphic order parameter for the condensation transition. To see this, one reforms $\text{J}_{\text{c}}(z,z^*) = |\text{J}_{\text{c}}|e^{\mathrm{i}S}$ so that the phase $S$ may have a pole at $z_{\mathrm{m}}$.  
The winding number thus follows, in a form of
\begin{equation}
    Q = \frac{1}{2\mathrm{\pi}} \bigointssss {\bf A} \cdot \text{d}{\bf R} = \frac{1}{4\pi} \bigintssss F_{\mu \nu}  \text{d}R_{\mu} \wedge \text{d}R_{\nu}
\label{WQ}
\end{equation}
where the connection tensor $F_{\mu \nu} = \partial_{\mu} A_{\nu} - \partial_{\nu} A_{\mu}$ and the curvature ${\bf A} = \nabla S$. $Q$ is thus a topological invariant in the fundamental group $\pi_1[\text{U}(1)]=\mathbb{Z}$. As shown in Fig.\ref{flux}(a,c), the phase of $\text{J}_{{\rm c}}$ always undergoes the same $2\pi$ rotation as the loop, when tracing along a closed loop encircling the peak and returning to the initial position. This leads to $Q=1$.


Table \ref{T1} collects the symmetry and graph topology for the three phases. Notably, the polariton condensation (PC) as a promising driven-dissipative phase obeys the rEOM sharing the structure of Eq.(\ref{EOMP}) \cite{Zhang_PRB2022,Lagoudakis_PRL2022}. The PC thus resides in the NCB regime, exhibiting a graph texture distinct from the BEC phase.


Fig.\ref{flux} shows the 2D curl fluxes given by Eq.(\ref{Jc}). From Fig.\ref{flux}(a), it turns out that in the above-threshold regime, $\text{J}_{\text{c}}=0$ around the peak of the number distribution $P_{n,N}$. $|{\rm J}_{{\rm c}}|$ further exhibits a summit-crater landscape, in Fig.\ref{flux}(d), revealing a ring ridge that locates an optimal curl flux which reflects the order parameter in Eq.(\ref{WQ}). These are align with Eq.(\ref{intJc}). In the below-threshold regime, however, the curl fluxes diminish. This is a thermal phase in a broader context, as depicted in Fig.\ref{flux}(b). Moreover, in Fig.\ref{flux}(a,c), ${\rm J}_{{\rm c}}$ shows a good agreement with the currents calculated from the curl network approach using Eq.(\ref{Jnet}).


\section{Analysis with curl flux network}

$J_{(n,N)\rightarrow (n',N')} \neq 0$ in Eq.(\ref{Jnet}) indicates a global nature on the current network. Because of the divergence-free nature of net currents on edge from Eq.(\ref{EOMP}), the current network can be decomposed into a series of closed loops. To confirm this, we calculate on the hexagonal-grid graph the current vectors that are formed by three net edge currents at site $(n,N)$. It shows loop feature surrounding the peak of $P_{n,N}$, i.e., in Fig.\ref{flux}(c), when above the pump threshold. This implies graph topology of a winding number $Q=1$, and is consistent with the results from continuous limit, i.e., Fig.\ref{flux}(a).

The network--composed by elementary triangles $\triangledown,\vartriangle$--can be reformed into cycle frequency along certain loop trajectories, i.e., $J_{(n,N)\rightarrow (n',N')} = \sum_C (f_{C_+} - f_{C_-})$ where $f_{C_{\pm}}$ is the cycle frequency along the trajectory $C$ [the summation is over all the trajectories involving the edge $(n,N)\rightarrow (n',N')$] (Appendix D). The cycle frequency quantifies the number of rounds which the system can transit through a complete cycle per unit time \cite{Ren_PRL2022,Schnakenberg_RMP1976,Hill_book,Esposito_JSM2014}. From the algebraic graph theory, we find the loop affinity $\Phi = \ln (f_{C_+}/f_{C_-})$ that follows the rule 
\begin{equation}
  \Phi = \sum_{\text{all}\ \triangledown_i\ \text{in}\ C} \Phi_{\triangledown_i} + \sum_{\text{all}\ \vartriangle_i\ \text{in}\ C} \Phi_{\vartriangle_i}
\label{Phitr}
\end{equation}
when along the trajectory $C$, where $\Phi_{\beta_i} = \ln \big(\Pi_{\beta_i^+}/\Pi_{\beta_i^-} \big)$ is the affinity on individual $\beta_i \in \{\triangledown_i, \vartriangle_i \}$ and $\Pi_{\beta_i^{\pm}}$ is a product of the transition rates along individual $\beta_i$ on the graph (Theorem 1 in Appendix C). $\Phi$ provides a measure of deviation from the equilibrium state, as known from the graph theory.

\begin{figure}[t]
\centering
\includegraphics[scale=0.22]{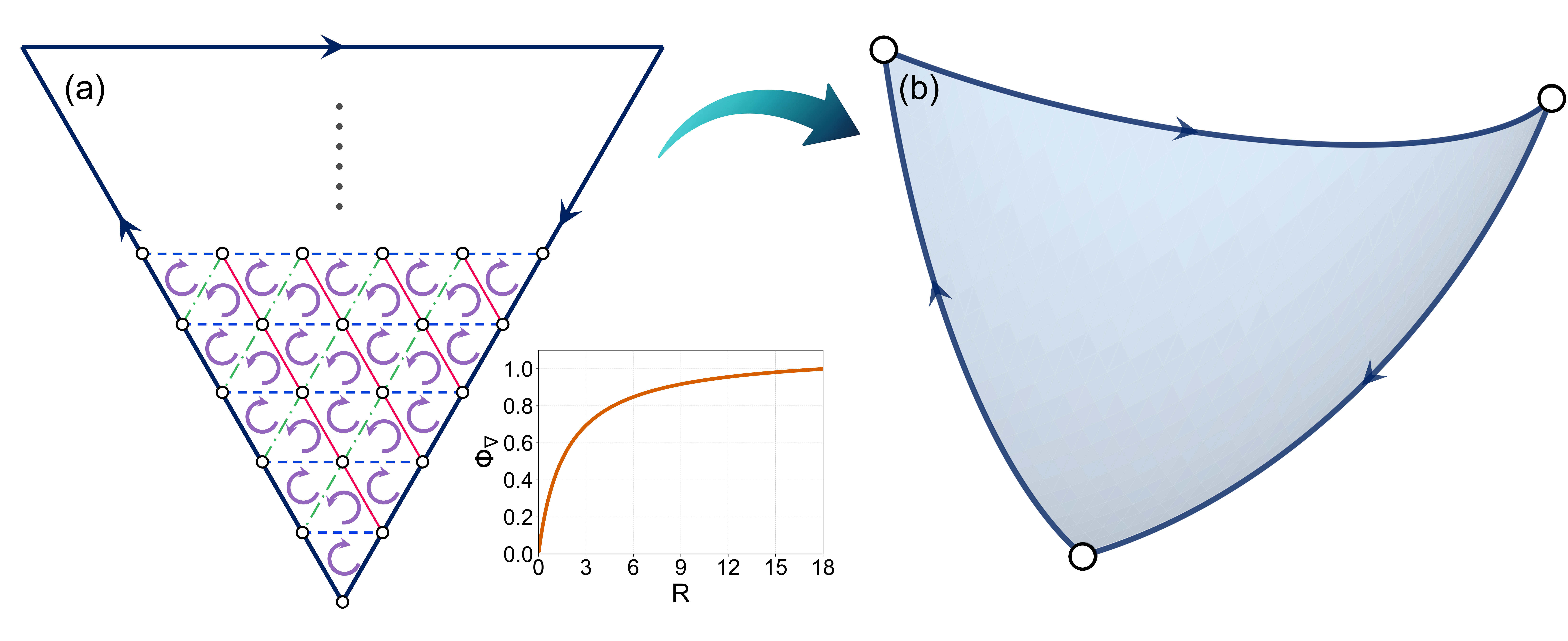}
\caption{Illustration of the loop affinity $\Phi$. (a) The affinities for $\triangledown,\vartriangle$ such that $\Phi_{\vartriangle} = -\Phi_{\triangledown}$ indicate opposite cycles (purple arrows). An edge flow is presented thereby (big arrows). Small panel: loop affinity vs. pump rate. (b) Loop affinity in Eq.(\ref{Ast}) is  regardless of deformations of the graph.}
\label{affinity}
\end{figure}

One further notes from Eq.(\ref{AF}) 
$\Pi_{\triangledown_i^+}/\Pi_{\triangledown_i^-}=\Pi_{\triangledown^+}/\Pi_{\triangledown^-} $ independent of the site index $(n,N)$, and  
$\Phi_{\triangledown}+\Phi_{\vartriangle}=0$ (Appendix C). 
It is convenient to rewrite $\Phi$ in terms of \emph{staggered area} on the graph. Defining $\varepsilon(T)=1$ for a $\triangledown$ plaquette and $\varepsilon(T)=-1$ for a $\vartriangle$ plaquette, the loop affinity in Eq.(\ref{Phitr}) further yields
\begin{equation}
  \Phi = {\cal A}_{\rm st}\Phi_{\triangledown}, \quad
  {\cal A}_{\mathrm{st}} = \sum_{T\subset R} \varepsilon(T) = m-n
\label{Ast}
\end{equation}
where ${\cal A}_{\mathrm{st}}$ measures the net area inside $C$; $m,n$ count the number of $\triangledown, \vartriangle$ inside $C$.

$\Phi$ scales with the staggered area determined by the bipartite coloring of the plaquettes. Therefore one can expect a  unidirectional flow along the outer edge of the graph, as depicted in Fig.\ref{affinity}.

It is worth noting from our model that the loop affinities associated with $\vartriangle, \triangledown$ are opposite. This means the graph topology with the PCs differs from the topological edge chiral modes that have been explored in electronic materials \cite{Zhang_RMP2011,Tang_PRX2021}. Nevertheless, the nonlocal curl flux emerges from the probability edge currents [i.e., Eq.(\ref{Jnet})] that may generate optimal loops, as proven in $\text{J}_{{\rm c}}(z,z^*)$ exhibiting the summit-crater landscape. Such a topological structure therefore dictates a dynamics distinct from the topology-protected chiral edge states.

\section{Discussion and summary}

The NCB phase has been observed extensively by measuring the intensity of emitted photons. Such a conventional technique is unable to access neither the counting statistics nor the fluctuations, which are however essential for understanding nonequilibrium properties of matter. Therefore new spectroscopic schemes may have to be proposed, e.g., using the delayed photon-coincidence counting (DPCC) that would be a feasible candidate.

The DPCC measures the Glauber’s multi-photon coherence, i.e., the function $g_2(\tau) = \langle \eta^{\dagger} \eta^{\dagger}(\tau) \eta(\tau) \eta\rangle$. In this vein, the signal $\sim {\cal M} g _2(\tau)$; the prefactor ${\cal M}$ arising from temporal gate parameters in detectors. It is anticipated that the NCB may lead to oscillations in the DPCC signal. This is consistent with the cycling nature of flux $\text{J}_{\rm c}$ that is likely to be coherent. In the infrared regime, the temporal gates require a duration $\sim 100$ps, which is achievable in laboratories.

Moreover, the THz time-domain spectroscopy emerges as an alternative and useful approach for probing the transient dynamics of the NCB of phonons, granting direct access to spectral fingerprints such as mode softening, linewidth narrowing and field-induced shifts. By studying the phase and amplitude of transmitted fields, one may reconstruct local field distribution and identify cyclic current patterns that signifies the NCB. Therefore combining THz excitation and pump-probe scheme would be beneficial for the temporal mapping of condensation dynamics and fluctuations.

Finally let us look into the experimental feasibility of achieving the NCB phase \cite{Lundholm_SD2015,Nardecchia_PRX2018,Zhang_PRL2019}. For a generic estimation, taking $R=S$ would be reasonable so that the threshold is $R_{\rm c}\approx \bar{n} + \frac{1}{M\alpha} + \frac{1}{M}$. For the THz vibrations, the decay $\gamma \sim 1$ns. Using $M=100, \alpha \sim 0.05, \bar{n}\sim 10$ (ambient temperature), one has the power to create the NCBs $p=10 M R_{\text{c}}\gamma\hbar \omega_{\text{v}}\sim 2.7\times 10^{-9}$W. With the area $A\sim 10 \mu$m$^2$ of pump field spot on the sample and the cross section of light scattering for resonance absorption at THz wavelength $\sigma \sim 10^{-15}$cm$^2$, the number of photons needed for a conisderable capture is $N = A/\sigma \sim 10^8$. Therefore the pump power is estimated to be $P = N p \sim 1$W.


In summary, our work presents a thorough study of the nonequilibrium condensate of bosons, in conjunction with the Fr\"ohlich coherence. Our quantum theory elucidated the detailed-balance-breaking nature through the curl flux network, revealing topological variation when driven far from equilibrium. A generic order parameter was identified for the NCB, showing graph topology beyond the symmetry-breaking paradigm of phase transitions. The results clearly demonstrated, on the graphs of the curl flux network, that the PCs rest in the regime of NCB. Understanding the nonequilibrium phases of matter will enrich the studies of complex systems coupled to external fields, thereby significantly advancing the frontier of the statistical thermodynamics at mesoscopic scale.

\vspace{0.15cm}

F. L. and C. S. contributed equally to this work.

\vspace{0.15cm}

\begin{acknowledgments}
We thank Sam Sung Ching Wong from City University of Hong Kong for the instructive discussions. Z.\ D.\ Z. and F.\ L.  gratefully acknowledge  the support of the Excellent Young Scientists Fund by National Science Foundation of China (No. 9240172), the General Fund by National Science Foundation of China (No. 12474364), and the National Science Foundation of China/RGC Collaborative Research Scheme (No. 9054901). Z.D.Z. and D.\ L. also acknowledge the financial support from the Guangdong Provincial Quantum Science Strategic Initiative (No. GDZX2205001) and the City University of Hong Kong through the RGMS grant (No. 9229137). X.W. is supported by the FYUST start-up grant, the National Natural Science Foundation of China (Grant No. 12505042), and the National Science Foundation of Zhejiang Province (Grant No. LQN25A050004). Z. L. and C. S. gratefully acknowledge the support of the National Science Foundation (No. DMR-2145256).
\end{acknowledgments}


\begin{thebibliography}{}
\bibitem{OF}
The equation of motion for the off-diagonal elements will be presented elsewhere.
\bibitem{ABKH}
The coefficients ${\cal A}_{n,N} = \sum_{j=1}^M \langle n_j+1\rangle_{n,N}$, ${\cal B}_{n,N} = \sum_{j=1}^M \langle n_j\rangle_{n,N}$, ${\cal K}_{n,N} = \sum_{j=1}^M (\bar{n}_{j0}+1) \langle n_j\rangle_{n,N}$, ${\cal H}_{n,N} = \sum_{j=1}^M \bar{n}_{j0} \langle n_j+1\rangle_{n,N}$ where $\langle \cdots \rangle_{n,N}$ denotes the mean, given $n,N$. $\bar{n}_{j0} = [e^{(\omega_j-\omega_0)/T} - 1]^{-1}$ is the Planck factor.
\bibitem{Abelian}
The space of circles is isomorphic to the Abelian group, as seen from the product operation $\phi_2\cdot \phi_1$ that is commutative. $e^{{\rm i}n \theta}$ forms an irreducible representation of the Abelian group where the integer $n$ counts the number of rounds for a periodic boundary condition.
\end{thebibliography}
\end{document}